\documentclass[letterpaper]{article}
\usepackage{natbib, alifeconf}
\usepackage{amsmath, amssymb}
\usepackage{url, hyperref, cleveref}
\usepackage{svg, booktabs}
\usepackage{float}
\usepackage{comment}
\usepackage{enumitem}

\title{Prebiotic Functional Programs: Endogenous Selection in an Artificial Chemistry}

\author{
    Devansh Vimal$^{1}$,
    Cole Mathis$^{1,2}$,
    Westley Weimer$^{3}$, \and
    Stephanie Forrest$^{1,4,5}$ \\
    \mbox{}\\
    $^1$Biodesign Institute, Arizona State University, Tempe AZ, USA\\
    $^2$School of Complex Adaptive Systems, Arizona State University, Tempe AZ, USA\\
    $^3$Electrical Engineering and Computer Science, University of Michigan, Ann Arbor MI, USA\\
    $^4$School of Computing and Augmented Intelligence, Arizona State University, Tempe AZ, USA\\
    $^5$Santa Fe Institute, Santa Fe NM, USA\\
    \texttt{cole.mathis@asu.edu}
}
\newcommand{\scc}{\ensuremath{\lambda n.\ \lambda a. \ \lambda b.\ a \ (n\ a\ b)}}
\newcommand{\add}{\ensuremath{\lambda n.\ \lambda m. \ n \texttt{ scc } m}}
\newcommand{\addtwo}{\ensuremath{\lambda n. \texttt{ scc } (\texttt{scc } n)}}
\newcommand{\tru}{\ensuremath{\lambda a.\ \lambda b. \ a}}
\newcommand{\fls}{\ensuremath{\lambda a.\ \lambda b. \ b}}
\newcommand{\addp}{\ensuremath{\lambda s.\ \lambda a. \ \lambda b.\ a\ s\ b}}
\newcommand{\scomb}{\ensuremath{\lambda x.\ \lambda y. \ \lambda z.\ x\ z\ (y\ z)}}
\newcommand{\kcomb}{\ensuremath{\lambda x.\ \lambda y. \ x}}
\newcommand{\icomb}{\ensuremath{\lambda x.\ x}}

\date{May 2025}

\begin{document}

\maketitle

\begin{abstract}
    Artificial chemistry simulations produce many intriguing emergent behaviors, but they are often difficult to steer or control.
    This paper proposes a method for steering the dynamics of a classic artificial chemistry model, known as AlChemy (Algorithmic Chemistry), which is based on untyped lambda calculus.
    Our approach leverages features that are endogenous to AlChemy without constructing an explicit external fitness function \textcolor{black}{or building learning into the dynamics}.
    We demonstrate the approach by synthesizing non-trivial lambda functions, such as Church addition and succession, from simple primitives.
    The results provide insight into the possibility of \textcolor{black}{endogenous selection in diverse} systems such as autocatalytic chemical networks and software systems.
\end{abstract}

\section{Introduction}

Living systems emerged from unconstrained abiotic mixtures of molecules~\citep{smith2016origin}.
The field of prebiotic chemistry aims to understand how nonspecific and reactive abiotic compounds self-organize into the restricted subsets of chemical space that are characteristic of life~\citep{muchowska2020nonenzymatic}.
Understanding and predicting chemical reactivity is a notoriously difficult problem, and even the best available models are only capable of simulating picoseconds to nanosecond time scales in detail~\citep{zhang2024exploring}.
Artificial chemistry models represent a coarse-grained approach that do not model the full detail of chemistry but still enable exploration of self-organization in reactive mixtures~\citep{banzhaf2015artificial}.
These models, like other artificial life (ALife) models, can exhibit emergent phenomena that are non-trivial consequences of the microscopic dynamics.
However, they are typically difficult to control or steer, as the relationship between model features and model behavior is, itself, complex.
This has contributed to significant debate as to whether the self-organized entities that emerge from these models can serve as a substrate for evolution and selection in a rigorous sense~\citep{vasas2010lack, vasas2012evolution, danger2020conditions}.

In contrast to artificial chemistry, evolutionary computation (EC) models rely on an externally defined, often static, fitness function to guide behavior, and they typically lack open-ended evolution 
%of many artificial chemistry (or ALife) systems
~\cite{miikkulainen2021biological}.
Although EC has been used to discover rules for ALife systems that lead to desired  emergent behaviors, e.g. ~\citep{Mitchell99EvolvingCA,hecker2015beyond}, and some ALife systems build adaptive mechanisms in to the system, e.g. ~\citep{tierra-1993,lenski1999genome}, ALife systems are typically not easily steerable once the underlying dynamics have been defined.
Here, we demonstrate how to guide
%a steering mechanism, or implicit selection, in 
a classic artificial chemistry model called AlChemy~\citep{fontana1994arrival,fontana1994would,fontana1996barrier} without altering its internal dynamics.
Our mechanism uses features endogenous to the dynamics of the model while allowing high-level specification of desired emergent outcomes.

AlChemy (short for Algorithmic Chemistry) is an artificial chemistry model that models key features of chemistry as a formal computational system known as the \textit{lambda calculus}.
%to capture key features of chemistry.
In brief, AlChemy represents molecules as \textit{lambda expressions}, which can be thought of as simple computer programs.
Pairs of these programs interact by using one program as input to another (called a collision), yielding a new program through their combination.
This process is analogous to chemical reactions among molecules.
We hypothesize that this approach can be extended to synthesize algorithmically useful programs via an implicit steering mechanism, which endogenously guides the bottom-up dynamics of the system to select target `molecules' (i.e., lambda expressions).
We achieve this by adding carefully-designed lambda functions to the AlChemy soup, which function as reagents for the simulation.
These functions (1) encode the desired behavior of the target molecules as unit tests~\citep{runeson2006survey}, and (2) output copies of  successful functions to the simulation.
Thus, the dynamics of AlChemy are guided endogenously by the properties of these functions.

A key contribution of the paper is the design of an effective \textit{amplifier} function.
Its presence in the simulation steers the dynamics such that (1) molecules that pass the tests (i.e., exhibit desired behavior) are amplified, while (2) the tests are persist in the population until desired molecules emerge.
This allows the system to select for the desired functional properties without specifying the precise program (`molecule') ahead of time.
Formally, an amplifier function is a second-order function that, when applied to a candidate lambda function, returns duplicates of the candidate function if it passes the test, but returns itself (the amplifier) if the test fails.
This structure has the effect of amplifying the abundance of functions that pass the embedded test (in the case of success) or preserving the test for future interactions (in the case of failure).
Critically, an amplifier is itself a lambda expression, and can thus be added to the  AlChemy state without further manipulation of the simulation.

This approach can be thought of as a simple form of program synthesis, in which the programmer specifies desired behavior via unit tests, and the AlChemy system produces a program that passes those tests.
It can also be thought of as an automated approach to chemical synthesis that monitors the system in real time, and when a key observable is detected (the unit tests are passed) the solution is purified to increase the concentration of the desired product~\citep{asche2021robotic}.
Alternatively, one can think of this approach as a version of Dynamic Combinatorial Chemistry (DCC) for functional programs. Although the reactions here are not thermodynamically controlled or strictly reversible, the design principle is the same: one designs an experiment that will amplify a desired target rather than designing the target itself~\citep{lehn2001dynamic}.
We demonstrate the effectiveness of our approach with test functions that successfully select and amplify arithmetic functions from an initial population of four primitive lambda calculus expressions.

\section{Background and Related Work}
\label{sec:background}

\paragraph{AlChemy}
AlChemy was introduced by Fontana and Buss in 1992 to study how higher-level organizations might spontaneously emerge from basic molecular interactions.
They refer to this as the \textit{existence problem} of population genetics~\citep{fontana1994would, fontana1994arrival}.
AlChemy abstracts the details of chemistry and uses lambda expressions to represent molecules.
Lambda expressions are computational functions written in the notation of the lambda calculus.
Function application represents molecular interaction, and lambda reduction (see below) models chemical reactions.
All functions are reduced to normal form using a process that is well-defined in the lambda calculus.
However, this reduction process is not guaranteed to terminate.
To avoid runaway reductions, AlChemy terminates the reaction after a fixed number of reduction operations or a maximum memory allocation. Here we use a threshold of 8000 reduction steps and a limit of 1000 vertices on the syntax tree represented by the expression.
The dynamics of an AlChemy simulation proceed as follows:
1) Generate a set of $N$ randomly generated lambda expressions, called a \textit{soup};
2) Select two expressions at random from the soup, call them $A$ and $B$;
3) Apply $A$ to $B$, and attempt to reduce to normal form within a constant number of reductions;
4) If (3) succeeds, add the newly generated expression to the soup and remove a different, randomly chosen, expression from the soup, and otherwise discard the irreducible expression (this guarantees a constant size population);
5) Iterate steps 2--4 $T$ times.
Each function application and attempted reduction is called a collision.
Time is measured by the number of collisions that have occurred.

Fontana \& Buss showed that in the long time limit ($T \gg N$), these dynamics can produce a set of mutually dependent lambda expressions, such that the application of any two expressions in the set yields another expression in the set~\citep{fontana1994would}. 
These sets are analogous to autocatalytic sets of molecules.
Fontana \& Buss identified a hierarchy of these collectively reproducing sets, which included selfish replicators, small hypercycles, and complex ecosystems~\citep{fontana1994arrival}.
We recently revisited the AlChemy simulation, focusing on the stability and diversity of the emergent organizations (autocatalytic networks) using the original code base~\citep{mathis2024self}.
Here, we show how to steer the AlChemy dynamics to produce useful programs, using a testing approach that relies on the functional nature of AlChemy.
% Pemma complains of an abrupt ending here

\paragraph{The Lambda Calculus}
The lambda calculus is a formal system of computation~\citep{church1985calculi}.
It specifies interactions between pure functions, i.e., functions that have no internal state.
We summarize it here; see \citep{barendregt1984lambda} for a rigorous treatment.
The lambda calculus operates on structures called lambda expressions defined over a finite alphabet $\Sigma$.
%SF deleted 'recursively-defined' because Item 1 is not recursive, so we might have to say mutually recursive, which didn't seem worth the trouble.
Expressions take one of three forms:
\begin{enumerate}[nosep, noitemsep]
    \item Single variables, $x$, chosen from $\Sigma$.
    \item Lambda abstractions of the form $\lambda x.\ E$, where $x$ is a variable, and $E$ is a lambda expression.
    This form describes a function with argument $x$, binding all other $x$-variables within its body $E$, which are not already bound to another abstraction.
    \item Applications of the form $(E_1\ E_2)$, where $E_1$ and $E_2$ are expressions.
    This form describes the composition of objects.
    If $E_1$ is a function, then $E_2$ is passed to $E_1$ as an argument.
\end{enumerate}
The lambda calculus also defines two transformation operations, $\beta$-reduction and $\alpha$-renaming, and a deterministic strategy for applying these operations to any lambda expression.
Both operations are rewriting rules that rewrite the given lambda expression into another lambda expression.
They can be iteratively applied until the lambda expression is in a form such that no rewrites are possible.
This is called a \textit{normal form}.
Once a lambda expression is in normal form, it has terminated its computation.
Because the lambda calculus is Turing complete, there exist lambda expressions that never reduce to a normal form---that is, the deterministic reduction process may never terminate.

The $\beta$-reduction operation takes as input expressions of the form $(\lambda x.\ E_1)\ E_2$.
The input expression is rewritten by substituting each copy of $x$ in $E_1$ with a copy of $E_2$.
The $\lambda x$ and $E_2$ in the original expression are then removed.
This operation is analogous to a function call in a modern programming language: $\lambda x. E_1$ is the function, and $E_2$ is the argument passed to the function.
The $\alpha$-renaming operation rewrites all variables within the scope of a lambda expression, substituting for each another variable chosen from $\Sigma$, unless a variable is already bound to another variable, in which case it is unchanged.
This operation is analogous to renaming a variable in an arithmetic expression.
For example, the $x$ variable in the expression $\lambda x.\ \lambda y.\ x$ can be $\alpha$-renamed to $z$, producing $\lambda z.\ \lambda y.\ z$.

If an expression is not in normal form, then the leftmost, outermost reducible expression is $\beta$-reduced first.
During a reduction, the substituting expression may have a variable that conflicts with an existing bound variable within the same scope (for example, the first $y$ within the reducible expression $(\lambda x. y\ x)\ y$).
In this case, we apply $\alpha$-renaming to the variable in the 
%FIXME: SF thought we agreed to change 'subtituent' to 'substituted'. -
substituted expression to preserve the semantic meaning of the expression before and after reduction.
Two expressions are said to be \emph{exactly isomorphic} if they have the same syntactic structure and the same semantic structure under $\alpha$-renaming.

\begin{table}[t]
    \centering
    \begin{tabular}{lp{0.25\textwidth}}
        \toprule
        \textbf{Function} & \textbf{Lambda Expression}          \\
        \midrule
        \verb|true|     & \tru                                  \\
        \verb|false|    & \fls                                  \\
        \verb|and|      & $\lambda a.\ \lambda b.\ a\ b \ a$    \\
        \verb|or|       & $\lambda a.\ \lambda b.\ a\ a \ b$    \\
        \verb|not|      & $\lambda a.\ a \texttt{ false true}$  \\
                                                                \\
        \verb|zero|     & \fls                                  \\
        \verb|one|      & $\lambda a.\ \lambda b.\ a\  b$       \\
        \verb|two|      & $\lambda a.\ \lambda b.\ a\ (a\  b)$  \\
        The integer $n$ & $\lambda a.\ \lambda b.\ (a\ (a\ \ldots (a\ b)))$, where $a$ is applied to $b$ $n$ times \\
                                                                \\
        \verb|S|        & \scomb                                \\
        \verb|K|        & \kcomb                                \\
        \verb|I| (identity function) & \icomb                   \\
        \verb|P|        & \addp                                 \\
                                                                \\
        \verb|scc|      & \scc                                  \\
        \verb|add|      & \add                                  \\
        \verb|add2|     & \addtwo                               \\
%        \verb|amplify-add| & $\texttt{atf} = \lambda f.\ (= (f\ 2\ 3)\ 5 )\ f\ \texttt{atf}$ \\
        \bottomrule
    \end{tabular}
    \caption{Key lambda expressions used in this work.}
    \label{tab:lambdas}
\end{table}

\paragraph{Lambda Notation}
Common functions in the lambda calculus used in this work are shown in Table~\ref{tab:lambdas}.
A lambda expression with no free (i.e., externally-defined) variables is said to be in closed form and is called a \emph{combinator}; all of the expressions in Table~\ref{tab:lambdas} are combinators.
The boolean literals \verb|true| ($\tru$) and \verb|false| ($\fls$) are functions that, by convention, take as input two arguments and return their first and second argument respectively.
Propositional connectives (i.e., \verb|and|, \verb|or|, \verb|not|) have the standard semantics.
Numbers are encoded using the standard Church numeral representation~\citep{barendregt1984lambda}.
A Church numeral $n$ is a function of two arguments.
It encodes $n$ applications of its first argument to its second argument.

Certain combinators are defined in the literature.
The $S$, $K$ and $I$ (identity function) combinators, taken together, are universal, i.e., they form a Turing-complete system, and any computation can be expressed as combinations of them.
We also describe the $P$ combinator, which permutes its arguments.
Our initial AlChemy population (\emph{soup}) consists only of copies of $S$, $K$, $I$ combinators, except
when we are amplifying addition, where we also include the $P$ combinator in the soup.
In this paper, we focus on evolving three functions that manipulate numbers.
Each function takes in one or more Church numerals and returns a single Church numeral:

\begin{enumerate}[nosep, noitemsep]
    \item \emph{Successor} (\verb|scc|), a function that increments its argument by 1: $\scc$
    \item \emph{Add-two} (\verb|add2|), a function that increments its argument by 2, by applying \verb|scc| twice: $\addtwo$
    \item \emph{Addition} (\verb|add|), a function that takes in two arguments, $n$ and $m$, and applies \verb|scc| to $m$ exactly $n$ times thus producing the sum $n + m$: $\add$.
\end{enumerate}

We focus on these functions because they encode non-trivial computation, 
%demonstrate a powerful structure, 
%SF made this change.
\textcolor{black}{have testable semantics, and show how the discovery of one function} (\verb|succ|) \textcolor{black}{can serve as a stepping stone to} \verb|add|.
For example, just as addition can be implemented as $n$ applications of successor $m$, so multiplication can be accomplished by applying \verb|add n| $m$ times to $0$: \verb|mul| = $\lambda n.\ \lambda m.\ m\ (\texttt{add~} n)\ 0$.

\paragraph{Related Work}
One approach to designing ALife models with desired properties is to use EC to evolve the low-level rules, as in Mitchell and Crutchfield's famous work on evolving cellular automata (CA) rules~\citep{Mitchell99EvolvingCA}.
This approach relies on an explicit fitness function (e.g., how many cells are in a given state) after some number of iterations of the CA, and the discovered rules determine the CA's dynamics.
Other approaches 
%are closer in spirit to our work, because they 
evolve a rule set or adjust weights while the system runs, e.g., learning classifier systems~\citep{jhh-induction-86}, Echo~\citep{Hraber1997TheEO}, and countless reinforcement learning algorithms dating back to \citep{Samuel1959,Sutton2005ReinforcementLA}.
In contrast, the dynamics of AlChemy do not explicitly include a learning step, relying instead only on deterministic interactions between lambda expressions.
Similarly Hillis's classic work on co-evolving parasites and sorting networks also relies on embedding an evolutionary dynamic directly in the system~\citep{hillis1990co}, \textcolor{black}{by introducing the idea of two competing populations, each evolving against the other in what is known as \textit{competitive co-evolution} \cite{competitive-co-evolution}.  Although its goal is quite different and the context is an EC with external fitness and genetic operators, \textit{lexicase selection}~\cite{spector2012assessment} resembles the amplifier function presented below. The algorithm selects parents for the next generation by assessing performance on a randomly chosen subset of test cases that define the problem domain.}
%\textcolor{black}{SF lexicase stuff, competitive co-evolution}.

Program synthesis discovers (synthesizes) programs that satisfy a stated specification~\citep{gulwani17synthesis}.
Traditional program synthesis often uses logical approaches (such as constraint solving or deduction) or search-based methods.
These have led to the successful synthesis of non-trivial program kernels (e.g., dynamic programming algorithms) from formal specifications~\citep{saurabh10synthesis}.
Program synthesis has long been goal in software engineering~\citep{manna71synthesis}, and it has enjoyed renewed interest with the advent of large language models~\citep{austin2021programsynthesislargelanguage}.
By contrast, our approach focuses on the use of amplifier test functions, which more closely mimics familiar unit tests~\citep{runeson2006survey}.

Several other ALife approaches synthesize or evolve program code.
Tierra and Avida explored the endogenous generation of programs, but differ from ours by relying on competition among individual programs and built-in self-reproduction~\citep{lenski1999genome} \textcolor{black}{making these systems more similar to artificial organisms, in contrast to the more distributed artificial chemistry system used here, where
autocatalytic reproduction emerges spontaneously from the dynamics of the system and  
%Although simple self-replication is possible, reproduction often occurs collectively through the interaction of many programs~\cite{mathis2024self}. }
%In our work, self-reproduction emerges from the dynamics of the system and 
the reagents that we add to the system~\citep{mathis2024self}}. Similarly,
~\cite{arcas2024computationallifewellformedselfreplicating} studies the rise of self-replicators in minimal programming languages but uses a Turing Machine based model rather than the functional approach we take here.
Finally, Genetic Programming and more recent variants use EC to generate software in the context of an external fitness function~\citep{koza1999genetic,weimer2010automatic}.

\section{Implementation}
\label{sec:impl}

One contribution of this paper is the reimplementation of Fontana and Buss's AlChemy system in a modern programming language (Rust)\footnote{\url{https://github.com/AgentElement/functional-supercollider}}.
The original implementation was written in an early version of C and has certain incompatibilities with modern compilers.
Our implementation supports more replicates of runs and larger simulations, e.g.,~ the runs in this paper contain up to $10^6$ expressions running for up to $10^8$ collision steps.
Small-scale experiments (100 soups or fewer) in this work were run on an AMD Ryzen 7840U CPU with 32 GiB of memory, and large-scale experiments were run on the ASU \textit{Sol} supercomputer~\citep{jennewein2023sol}.  Run parameters were determined from preliminary experiments and the results reported in~\citep{mathis2024self}.

\section{Amplifier Functions}
\label{sec:amplifier}

The cornerstone of our approach are \textit{amplifier test functions}: generic second-order functions that duplicate functions that pass a series of unit tests.
%FIXME-Devansh: This is a place where the confusion exists---if the amplifier function encodes a single unit test, then doesn't the test function get copied if it passes that one.
%FIXME: we should say here what a unit test is, i.e., a set of inputs together with the desired output.  This is especially confusing because coming up in a few sentences you talk about the amplification factor, which is different.
Given a unit test, the amplifier function wraps it such that application of the amplifier to a function $f$ will return copies of $f$ if $f$ passes the test. Otherwise, if the test fails, the amplifier will return itself.
%SF deleted next sentence because it calls attention to the fact that there is a 'programmer' and we havent' established that writing unit tests is easier than writing the function itself.
%This separation of concerns allows the programmer to focus on writing the unit tests  (i.e., on encoding desired behavior) while the generic amplifier structure handles replication for any such test.
We refer to the number of copies of $f$ as the \textit{amplification factor}.
In the success case, the amplifier test amplifies the number of good functions in the population.
In the failure case, by returning itself, the amplifier test maintains its own population.

We illustrate the amplifier design with an example.
Suppose our goal is to produce a soup that contains a significant proportion of addition functions, e.g., enough for easy detection.
The programmer creates a unit test for addition, called \texttt{test}, such as \verb|(= (f 2 3) 5)|.
An amplifier test function \texttt{atf} for that \texttt{test} is:
\[
    \texttt{atf} = \lambda f.\ (\texttt{test } f)\ f \ \texttt{atf}
\]
This amplifier test function returns itself (\texttt{atf}) if $(\texttt{test } f)$ is false, and returns its argument ($f$) if $(\texttt{test } f)$ is true.
In this example, only one copy of $f$ is returned (the amplification factor is 1).
\footnote{As an explicit example, we obtain the amplifier test function $ \texttt{atf} = \lambda f.\ (= (f\ 2\ 3)\ 5 )\ f\ \texttt{atf} $, upon substituting the expression $(= (f\ 2\ 3)\ 5)$ for \texttt{test}}.
Note that the amplifier test functions do \emph{not} generate the desirable function.
Rather, they amplify desirable functions when they appear in the soup and collide with the amplifier function.
%SF: We've beaten this horse to death by now. 
%i.e., that were generated by the native AlChemy dynamics.

In AlChemy, amplifier functions are mostly inert because they are large---often the largest expressions in the soup.
When used as arguments to other functions, they frequently fail to reduce to normal form.
The pragmatic reduction of AlChemy (reduce up to $T$ steps, and if the expression is still reducible or exceeds a certain memory limit, then fail the reaction) means that most collisions fail if the test function is used as the argument.
As a time-saving optimization, we can eagerly fail collisions that have amplifier test functions as arguments if we know they will not reduce in $T$ steps.

Amplifier test functions with high amplification factors can  amplify very rare spontaneous events. For example, the spontaneous evolution of addition from a soup of only $S$, $K$, $I$ and $P$ combinators is a rare event.
In our experiments, an amplification factor of 100 suffices to ensure that a successful addition function remains in the soup available for more collisions with test functions.
However, high amplification factors can lead to the proliferation and eventual domination of unwanted \textit{trickster functions}: functions that pass the test cases without actually behaving like the desired function (see Section \ref{sec:trickster}).
%We discuss trickster functions further in the later sections.

\section{Experiment: Amplifier Functions}
\label{sec:exp-amplifier}

The first experiment demonstrates that adding amplifier functions to soups does amplify desirable functions, as expected.  
The probability of an amplifier function colliding with a desired function is proportional to the concentrations of amplifier functions and target functions in the soup.
If there are too few amplifier functions, then they are unlikely to randomly collide with the desired function.
Thus, the initial emergence of desirable functions is fragile because good functions can collide with other functions, transforming them into undesirable functions, just as desired molecular products might react with other molecules to form undesirable products.

In this initial experiment, the target function is not expected to emerge spontaneously. Instead we initialize the soup with small quantities of the target function (between 0.002\% and 10\%) and the experiment asks whether the amplifier test functions actually amplify a target in the face of competing background reactions.  
For these experiments, the successor function is the target.

%We consider multiple soups,
% SF first tried to compress, but realized that this text is redundant with the paragraph that starts Figure 1.
%Each soup contains 5000 functions initialized with universal combinators (in equal proportions), and we vary both the initial concentration of amplifier functions and of target functions (successor).  For each tested combination, we consider 100 soups.
We measure the effectiveness of the amplifier tests by measuring the quantity of functions in the soup that are exactly isomorphic to a pre-written implementation of the desired function.
For example, to measure the effectiveness of the successor test function, we measure the concentration of functions in the soup that are exactly isomorphic to the expression $\scc$.
Because this is not the only possible implementation of successor, this measurement is conservative and thus provides a lower bound on the effectiveness of amplification.
Since it is undecidable to determine if two lambda expressions exhibit identical behavior on all inputs, an approximation of this kind is necessary.

\begin{figure}[t]
    \centering
    \includegraphics[width=\linewidth]{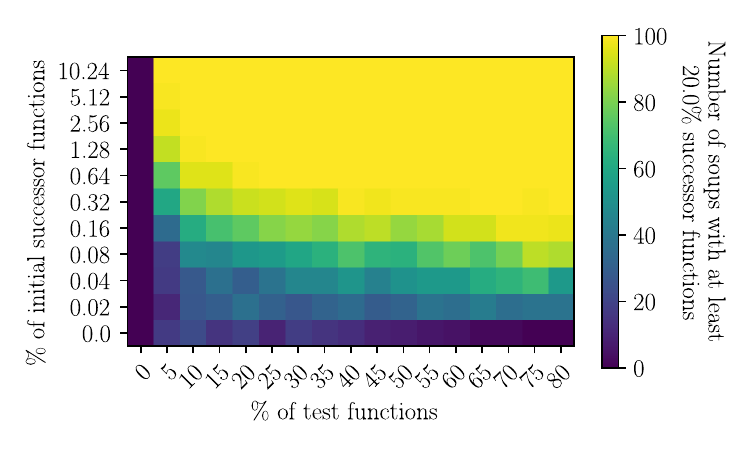}
    \vspace{-2.5em}
    \caption{
        Amplification of the successor function under varying initial conditions.
        For each pixel in the heatmap, 100 soups containing 5000 expressions are run for $10^6$ collisions.
        We measure the number of soups containing at least $20\%$ successor functions at the end of their runs.
        Each soup is initialized with a small fraction of successor functions ($y$-axis), and a large fraction of test functions ($x$-axis).
        Bright yellow indicates that all soups have a large quantity of successor functions, dark blue indicates that none do.
    }
    \label{fig:heatmap}
\end{figure}

Figure~\ref{fig:heatmap} plots the results as a function of both the initial number of successor functions and the percentage of test functions.
The horizontal axis plots the prevalence of test functions (as a fraction of the population) as the soup evolves and the vertical axis shows how many target functions (successor) were artificially introduced into the soup (as a fraction of the population).
The color indicates the percentage of soups with at least 20\% successor functions (out of 100 soups run).
We found that introducing as few as 5\% amplifier functions substantially amplified the presence of the successor function in the soup.
On the other hand, when the amplifier test functions were not included, no runs achieved a population of successor functions that exceeded 20\%.

The amplification of good functions is largely independent of the fraction of test functions introduced into the soup.
Desirable functions are more likely to be amplified by the test functions than undesirable functions, even if there are only a small number of them present in the system.
If the initial population contains more than $0.16\%$ successor functions and at least $10\%$ amplifier test functions, then after $10^6$ collisions, the population of successor functions exceeds $20\%$ in one half of the runs.
However, without the amplifier test functions there is no amplification and no soup developed a population of successor functions above $20\%$.
%Thus our method is limited by the spontaneous emergence of a function with the set of desired properties.
Thus, once a target function appears in the soup, the amplifier test succeeds and produces enough copies to facilitate easy detection.
%the soup becomes dominated by copies of that function.

\section{Experiments: Targeted Selection}
\label{sec:exp-targeted}

%FIXME: we seem to use target function and desirable function interchangably---sf opted to standardize on target in this section, for consistency and clarity.
Having established that a target function can be amplified, we next investigate whether the dynamics of AlChemy can generate target functions endogenously.
We focus on generating and amplifying target functions from soups that do not initially contain the target.
Some of these functions are building blocks that can be used to synthesize the final result, e.g.,~ the successor function is a building block for the addition function.  In the following, we study this building block and two other target functions: addition and add-two.
The successor function is a key building block for many arithmetic functions, simple add-two is an example of a function that very rarely emerges spontaneously, and addition is a non-trivial extension of successor that almost never emerges spontaneously.

In all three experiments, we measure the population of target and amplifier functions every 1000 collisions.
We perturb the soup with new amplifier functions every $10^5$ collisions to replenish their supply.
This is necessary because the dynamics of AlChemy randomly remove a function from the soup after each successful collision, thus diluting amplifier functions as the simulation proceeds.
Each run lasts $10^6$ collisions, and initially contains between 5000 and 6000 expressions, depending on the number of unit tests.
These parameters were chosen to match our available computational resources.

\begin{figure}[t]
    \centering
    \includegraphics[width=\linewidth]{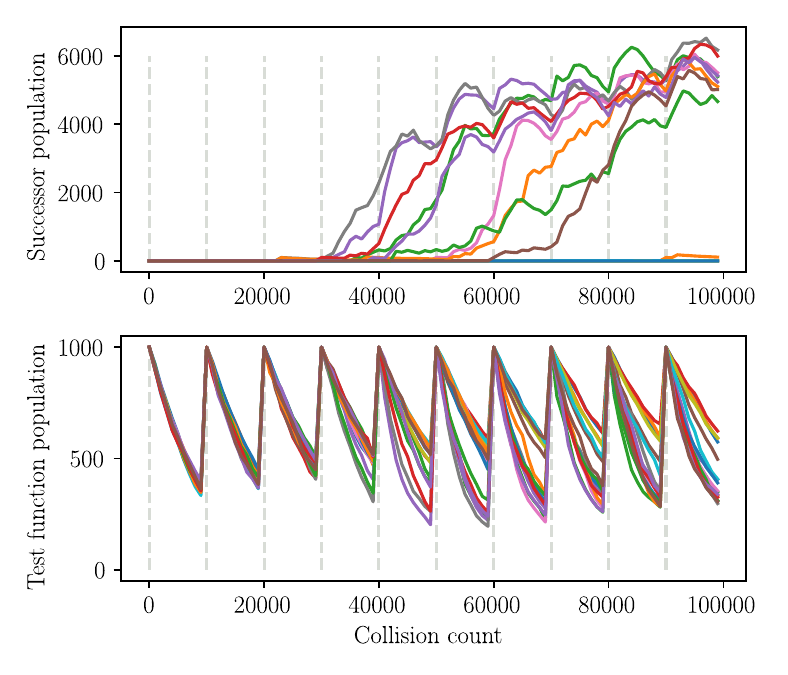}
    \vspace{-2.5em}
    \caption{
        Sixteen representative runs testing for and amplifying \textbf{successor}.
        (Top) Population of \texttt{scc} in the soup over time. Different colors correspond to different experimental runs with the same initial conditions.
        (Bottom)  The population of test functions over time.
        Every $10^5$ collisions (gray dashed lines), we introduce new amplifier functions to replenish those lost by random deletion and replacement.
        Each run is seeded with universal combinators and amplifier functions only.
    }
    \label{fig:scc-figure}
\end{figure}

\paragraph{Successor}
In the first small-scale experiment, we show that successor emerges reliably from a seed mixture of $S$, $K$ and $I$ expressions in equal quantities, plus amplifiers for successor.
The amplifier functions each take a function as an argument, apply the function to a fixed integer $0 \leq n \leq 20$, and check the result for equality with its successor: $n' = n + 1$.
These amplifiers take the form $\lambda f.\ (=\ (f\ n)\ n')$.
The $(=\ (f\ n)\ n')$ term acts as a test case.
Figure~\ref{fig:scc-figure}, shows the populations of \verb|scc| and amplifiers over time for sixteen runs.
Each run begins with an initial population of $6000$ functions (5000 seeds and 1000 amplifiers).

\begin{figure}[t]
    \centering
    \includegraphics[width=\linewidth]{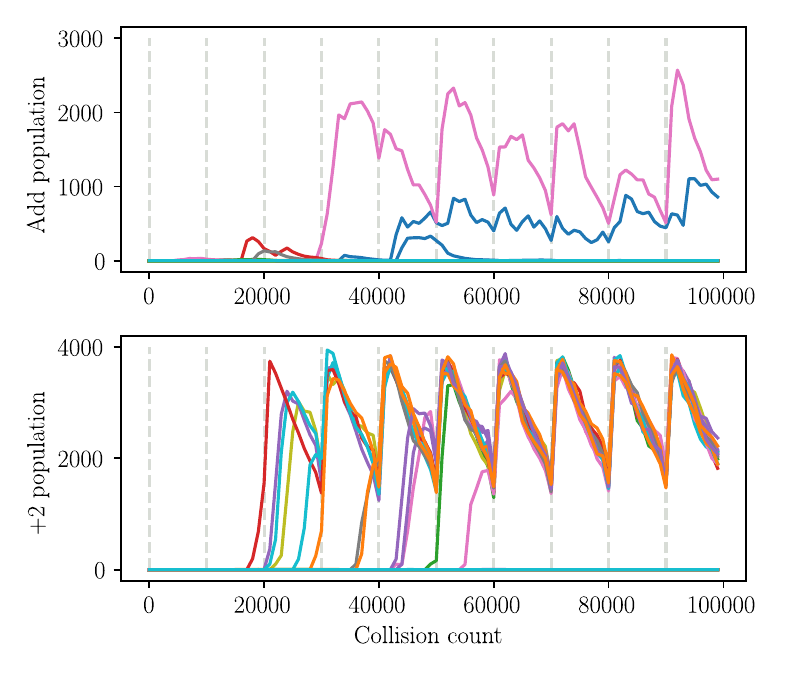}
    \vspace{-2.5em}
    \caption{
        $1000$ representative runs targeting \textbf{addition} and \textbf{add-two}.
        (Top) Population of \texttt{add} in the soup over time.
        (Bottom) Population of \texttt{add2} in the soup over time.
        Each set of runs was seeded with combinators and amplifier functions only.
        Different colors correspond to different experimental runs with identical initial conditions.
        The population of test functions over time is plotted in each panel.
        Every $10^5$ collisions (gray dashed lines), amplifier functions are added to replenish those lost by random deletion and replacement.
        Each run is seeded with universal combinators, \texttt{P}, and amplifier functions.
    }
    \label{fig:add-figure}
\end{figure}

\paragraph{Add-two}
Next, in a larger experiment of 1000 runs, we show that add-two emerges reliably.
The initial condition for this experiment is a soup that contains a seed mixture of $S$, $K$, and $I$ expressions in equal quantities plus amplifiers for add-two.
The amplifiers for add-two take a function as an argument, apply this function to a fixed integer $0 \leq n \leq 20$, and check the result for equality with $n' = n + 2$.
In the top panel of Figure~\ref{fig:add-figure}, we plot the population of \verb|add2| over time for 1000 runs. Each run begins with an initial population of $5000$ functions (15\% are amplifier functions, the remainder are seed functions).

\paragraph{Addition}
Finally, we investigate the emergence of addition in an experiment of 1000 runs.
The initial condition is a seed mixture of $S$, $K$, $I$ and $P$ expressions in equal quantities, plus an equal quantity of two amplifiers, one for addition and one for successor.
The amplifiers for addition each take a function as argument, apply this function to a fixed pair of integers $0 \leq n, m \leq 20$, and check the result for equality with their sum $\sigma = n + m$.
The tests take the form $\lambda f.\ (=\ (f\ n\ m)\ \sigma)$.
In the bottom panel of Figure~\ref{fig:add-figure}, we plot the population of \verb|add| over time for 1000 runs.
Each run begins with an initial population of $5000$ functions (15\% amplifier functions, and the rest seed functions).

In principle, the $P$ combinator is not strictly necessary for the evolution of addition, as the $S$, $K$, and $I$ expressions are universal.
However, in another experiment with 1000 soups seeded only with $S$, $K$, and $I$ expressions, we observed the spontaneous emergence of the addition function only once (not shown).
By random chance, the function was not amplified and diluted out of the soup.
Using the $P$ combinator increases the likelihood of generating addition because it produces addition when composed with $\verb|scc|$.
We hypothesize that the $P$ combinator would not be needed in larger-scale
versions of these experiments, an experiment we leave for future work. 
Including $P$ accounts for the fact that many more soups yielded addition in this experiment than add-two in the previous experiment.

\paragraph{Sensitivity}
We also evaluate three different initial populations, and test each population for the spontaneous emergence of the add and successor function, with and without tests.
This set of experiments is summarized in {Table~\ref{fig:tab}}.
We find that random inputs never generate good functions, unless the amplifier test is added and even then, rarely and only for the successor function.

\begin{figure}
    \centering
    \includegraphics[width=0.9
\linewidth]{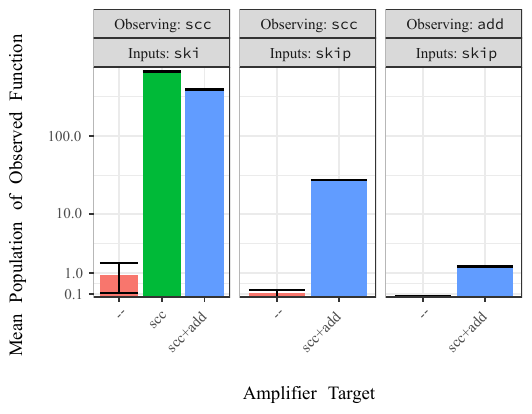}
    %\vspace{-1.5em}
    \caption{
        Discovery of numeric functions under different initial conditions. 
        Each panel shows a unique combination of initial inputs and target functions. 
        The horizontal axis indicates which test functions were included in the amplifier set. 
        The \textit{Inputs} row specifies the functions seeded into the soup at the start (either random functions, \texttt{SKI}, or \texttt{SKIP}). 
        Bar height represents the time-averaged population of the function indicated at the top of the panel. 
        Values are averaged across $10^3$ independent runs. ``--'' denotes cases where no amplifier functions were added.
        Cases with \textit{random} input functions are not shown as those yielded mean populations of less than $10^{-5}$.
}
    \label{fig:bar}
\end{figure}

If we seed the soup with expressions that are randomly generated by the binary tree method of~\citep{mathis2024self} instead of using $S$, $K$, $I$ then the probability of evolving any integer function is too small to reliably measure.
Across over 1000 runs, the successor function evolved fewer than five times from random expressions.
This suggests that randomly-generated expressions may either derail the spontaneous emergence of successor, or produce more tricksters that dominate the soup.

% \begin{table}[t]
%     \centering
%     \begin{tabular}{ccc}
%         \toprule
%         \textbf{Input Population} & \textbf{Amplifier Functions} & \textbf{Discovered}  \\
%         \midrule
%         Random  & --                        & --                                        \\
%         SKI     & --                        & $\texttt{scc}^*$                          \\
%         SKIP    & --                        & $\texttt{scc}^*$                          \\
%         Random  & \verb|scc|                & \verb|scc|                                \\
%         SKI     & \verb|scc|                & \verb|scc|                                \\
%         SKIP    & \verb|scc|                & \verb|scc|                                \\
%         Random  & \verb|scc|, \verb|add|    & \verb|scc|                                \\
%         SKI     & \verb|scc|, \verb|add|    & \verb|scc|, $\texttt{add}^*$              \\
%         SKIP    & \verb|scc|, \verb|add|    & \verb|scc|, \verb|add|                    \\
%         \bottomrule
%     \end{tabular}
%     \caption{
%         Discovery of numeric functions under various initial conditions.
%         The `Input Population' column names the set of expressions that each soup is initially seeded with.
%         `Amplifier functions' describes testing for the functions.
%         The `Discovered' column shows the functions seen in the soup.
%         A $*$ indicates that a very small quantity of functions was measured.
%     }
%     \label{tab:discovery}
% \end{table}

\section{Trickster Functions}
\label{sec:trickster}
The amplifier functions sometimes amplify a function that displays partially correct behavior but does not match the developer's intent.
Because the function passes at least one test, and because each amplifier function encodes exactly one unit test, it may be amplified and persist in the simulation.
We call such functions \textit{trickster functions}.
Of course, the complete set of unit tests provided to the system might also be inadequate to test exhaustively for the desired functionality in the target function.
Writing a complete test suite is a classic problem in software engineering and programming languages.
While testing can exhaustively check behavior for functions with small finite domains (e.g., \verb|and| or \verb|or|), in the general case of unbounded inputs or data structures, no finite test suite will suffice, as is the case for the functions explored here.

We developed three heuristics for mitigating trickster functions:

\begin{enumerate}[nosep, noitemsep]

\item When the test applies $k$ arguments to the candidate function, pass only those with at least $k$ leading abstractions (i.e., have arity $k$ or greater).

\item When the test applies $k$ arguments to the candidate function, pass only functions with bodies that reference those first $k$ arguments (i.e., fail functions that do not actually use their arguments).

\item Discard functions that are structurally identical to \verb|true| or \verb|false| wrapped in extra lambda abstractions.

\end{enumerate}

These heuristics are not universally applicable---for example, \#3 would not be appropriate if we were trying to find a function of three arguments that ignores those three arguments and always returns zero---but in practice, we found them to be efficient, \textcolor{black}{and applied them to all experiments reported here}.

\section{Discussion and Conclusion}

\label{sec:discussion}
The amplifier test function amplifies functions that might otherwise be fragile and unlikely to persist in the population, enabling implicit selection for functionality in AlChemy.
Such functions can be building blocks for synthesizing a desired final result or the final result itself.
Most functions that do not replicate themselves are fragile, including those required for many theoretical models, including  Turing Machines and Peano Arithmetic)~\citep{mathis2024self}.
For example, addition function returns a number and not a copy of the addition function, while a sorting function returns a sorted list and not a copy of itself.
Indeed, small functions that exactly replicate themselves (called \emph{quines} in the literature), are notoriously rare and hard to construct.

The amplifier design introduces implicit selective pressure into the bottom-up artificial life simulation dynamics in a manner that is analogous to the approach in \textit{dynamic combinatorial chemistry} (DCC)~\citep{lehn2001dynamic}.
In DCC, as in our system, a dynamic library of molecules is generated that rapidly inter-convert.
Key functionality can be selected out of this system by introducing a protein (analogous to our amplifier) that selectively binds to desirable functional groups (analogous to our unit tests).  This binding shifts the equilibrium of the internal dynamics toward the formation of the bound molecule, leading to amplification.

\paragraph{Limitations}
%Our test design increases the concentration of functions already present in the soup of expressions, thus requiring the spontaneous generation of desirable expressions.
We propose an \emph{amplification} solution, not a \emph{generation} solution.
Indeed, a key difficulty in amplifying the `add` function was the fact that it rarely emerged spontaneously from our initial conditions.
Additionally, AlChemy, as implemented here, uses the untyped lambda calculus.
Some synthesis tasks are likely to be more efficient in the typed lambda calculus.
Such an extension would limit collisions to occur only between compatible elements, and are comparable to orthogonal components in synthetic approaches, e.g., with Click Chemistry~\citep{paioti2024click}.

The emergence and selection of trickster functions in our system is unsurprising from a software engineering perspective, but it provides an interesting correspondence to synthetic chemistry.
The key challenge in synthetic chemistry is identifying reagents and reaction conditions that enable the chemist to produce some desired products (and hopefully many different ones) while preventing the system from producing undesirable side-products.
This is difficult because reactivity is fundamentally determined by the tendency (or not) of electrons to transition between orbitals. Making some orbitals accessible while preventing all others requires conditions that are exquisitely fine-tuned~\citep{szabo1996modern}.
In the case of our amplifier functions, we need to fine tune the tests using ever more diverse and precise test suites, just as chemists need to fine tune reaction conditions.
%Through this analogy with synthetic chemistry, 
We expect the spontaneous emergence of more complex functions in AlChemy to be rare for the same reason that we expect total synthesis of complex organics to be difficult: the desired function itself will be more rare, and the proportion of functions that are nearly identical to the desired function will grow as the complexity of the function grows~\citep{marshall2021identifying}.
Identifying approaches to overcome this challenge could provide insights for both program synthesis and prebiotic chemistry.

\paragraph{Future directions}
Although the amplifier described here was constructed using only the native lambda calculus representation of AlChemy,
the design may generalize to other representations.
In particular, we are interested in using a typed representation in an AlChemy-like system based on a functional programming language, such as Haskell, OCaml or TypeScript.
This would allow us to investigate whether a bottom-up system such as AlChemy can generate useful functions, and compare those implementations to large human coded standard libraries.
For example, if the system were initialized with existing functions, would other interesting functions emerge?
Ultimately, such an approach could provide a new approach to program synthesis, an active area of research in computer science~\citep{gulwani17synthesis}.
\textcolor{black}{Our experiments focused on functions that emerge spontaneously in a constant size soup through combination and reduction processes in AlChemy. An interesting question is how the size of the soup interacts with the amplifier functions, and how the length of the simulation affects results. We expect that larger system sizes would enable more distinct expressions to exist in the system at any given time, possibly accelerating the rate of discovery. 
However, the rate of amplification might change with system size, suggesting an interplay between efficiency and diversity. We suspect the broad features of this trade-off may be amenable to a mean-field analysis.}

The AlChemy system is an exciting approach to artificial chemistry because it retains some essential aspects of real chemical systems~\citep{mathis2024self, fontana1996barrier}.
Using the approach outlined here we can now explore how the system responds to selective pressures that can be imposed through an endogenous mechanism.
Our approach may enable future artificial chemistry approaches to more closely mirror the experimental procedures of modern prebiotic chemistry~\citep{surman2019environmental}.
We plan to further investigate how AlChemy can be used to make predictions about long-term chemical experiments and guide experimental design by developing the correspondence between AlChemy and real reaction platforms~\citep{asche2021robotic}.

\paragraph{Conclusion}
\textcolor{black}{This work addresses a long-standing limitation of conventional evolutionary algorithms, which typically rely on externally defined objective functions that return a numerical fitness value, or small set of values, for every individual in the population. 
This limitation prevents evolutionary computation from achieving the generative open-ended evolution that we observe in the natural world~\cite{miikkulainen2021biological}.}
\textcolor{black}{Similarly, prebiotic chemical experiments often exhibit diverse forms of self-organization, but open-ended evolution of those forms is limited because they lack genetic systems \citep{szathmary2006origin}.} 
This work provides insights into how more elaborate molecules might be produced through the step-wise selection of simpler molecules, if those simpler molecules can be amplified or made to persist longer against background reactions~\citep{asche2021robotic,surman2019environmental}
Ultimately, we aspire to develop systems that combine the self-organizing features of artificial chemistry, with EC to drive the emergence of novel functionality in chemical and software systems.

\section*{\textcolor{black}{Acknowledgments}}
We thank Kirtus Leyba, Pemma Reiter, Joe Renzullo, and the anonymous reviewers for their careful corrections and thoughtful suggestions.
SF gratefully acknowledges the partial support of NSF (CCF CCF2211750), DARPA (N6600120C4020), ARPA-H (SP4701-23-C-0074), and the Santa Fe Institute.

\footnotesize
\bibliographystyle{apalike}
\bibliography{ref}

\begin{thebibliography}{}
\bibitem[Fontana and Buss, 1994a]{fontana1994would}
Fontana, W. and Buss, L.~W. (1994a).
\newblock What would be conserved if" the tape were played twice"?
\newblock {\em Proceedings of the National Academy of Sciences}, 91(2):757--761.
\end{thebibliography}

\end{document}